# On the Lapse Function of FLRW Accelerated Expanding Universe in dRGT Massive Gravity Theory

Husin Alatas[1,3,#], Ahmad K. Falah[2], Trio Wibowo[1], Muhammad A. Qohhar[1], Bobby E. Gunara[2,3]

[1] Theoretical Physics Division, Department of Physics, Bogor Agricultural University, Jl. Meranti, Kampus IPB Darmaga, Bogor 16680, Indonesia
[2] Theoretical High Energy Physics and Instrumentation, Faculty of Mathematics and Natural Sciences, Institut Teknologi Bandung, Jl. Ganesha 10, Bandung 40132, Indonesia
[3] Indonesia Center for Theoretical and Mathematical Physics (ICTMP), Jl. Ganesha 10, Bandung 40132, Indonesia
[#] Corresponding author: alatas@apps.ipb.ac.id

**Abstract**

In this report, we discuss the behavior of coordinate-time dependent lapse function of FLRW metric of an accelerated expanding universe in the de Rham-Gabadadze-Tolley massive gravity theory. We find for the conventional dRGT formalism the corresponding lapse function can exhibit unphysical behaviors in the associated parameter space with a negative cosmological-constant-like term that leads to the decelerated universe model. To solve this problem, we introduce the so-called cosmological background density parameter to the perfect fluid stress-energy tensor which induces negative pressure. It turns out that this setup could overcome the existence of singular and negative square lapse function in the related parameter space and restore all the parameter space to admit only the accelerated expanding universe model.

**Keywords:** Lapse function; massive gravity; accelerated expanding universe

## 1. Introduction

Recently, the studies on accelerated universe expansion have led to the conjecture on the existence of dark energy, which suggests the relationship between the vacuum energy and cosmological constant [1]. However, this was found to suffer due to the extremely small value of the present time observed cosmological constant value. One of the possible alternative candidates to overcome this problem is the assumption that the gravitational field has to be mediated by a massive graviton [2], which is responsible for the accelerated expansion of the universe.

Among some recent suggested massive gravity theories, the de-Rham Gabadadze Tolley (dRGT) theory of spin-2 massive gravity has offered a well-behaved model which is relatively free from anomalies such as vDVZ discontinuity and Boulware-Desser ghost [3]. This massive gravity theory admits a self-accelerating universe model. In the meantime, the recent observation on the neutron stars collision originated gravitational and electromagnetic wave has revealed that there is a small propagation speed fractional difference between the graviton and photon of order $10^{-16}$ to $10^{-15}$ [4]. Though likely there is a possibility that this small difference can be neglected, but it is still reasonable to think that there is also an open possibility that the graviton may be massive [5].

In many reports on the cosmological applications of the dRGT theory e.g. Friedmann-Lemaitre-Robertson-Walker (FLRW) universe, it is common to consider at least the following two situations: (i) the corresponding stress-energy tensor is assumed to be in the form of perfect fluid and (ii) the consideration of coordinate-time dependent lapse function $N(t)$ in the corresponding dynamical metric [6-11]. However, we note that the behavior of lapse function, which acts as a gauge function, was rarely discussed in the literature. Here, we use the term "coordinate-time" to distinguish global time coordinate from the term "proper-time" that of related to a time measured by a normal observer [12].

This gauge-like function has been introduced in many reports only to derive the first Friedmann-Lemaitre equation based on effective Lagrangian method namely by varying the related action with respect to this function, without further investigation on its dynamical characteristics [6-11]. In standard cosmological models, this function is always set to trivial condition $N=1$ since we could always employ a time redefinition from coordinate-time to proper-time namely by replacing $N \equiv N(t)$ with $N=1$ in the corresponding metric.

On the other hand, for the dRGT theory, such lapse function of the dynamical metric cannot be automatically set to $N=1$ due to the existence of the fiducial metric lapse function that introduced to maintain the diffeomorphism invariance of the theory [3]. We demonstrate that the existence of such non-trivial $N(t)$ gives a new perspective on the cosmological application where the form of scale factor fixes the lapse function.

In this report, we show that in a specific homogeneous and isotropic flat universe model with FLRW on FLRW metrics, in which both dynamical and fiducial metrics are of FLRW metric forms, the considered lapse function might contain singularity and $N^2 < 0$ conditions in certain regions of parameter space, which leads to the unphysical condition. To solve this problem, we modify the corresponding stress-energy tensor by introducing a new constant parameter $\rho_0$ to the perfect fluid stress-energy tensor, that can be interpreted as of what we call "cosmological background density" parameter, which is related to the negative pressure similar to the cosmological constant. This choice of stress-energy does not alter the dynamical equations for matter and radiation density parameters. Moreover, in this proposal, we discuss a simple example of a flat universe model with the exponential scale factor.

We organize this report as follows. In section 2 we give briefly the formulation of the FLRW universe model including the definition of the related lapse function in dRGT theory, while in section 3 we discuss an example of flat FLRW accelerated expanding universe with the exponential scale factor. In section 4 we discuss the possible range of the corresponding newly introduce parameter. A conclusion is given in section 5. For the clarity of our discussion, we also include an appendix.

## 2. Lapse function in dRGT theory for FLRW universe

To discuss the corresponding self-accelerating universe model based on dRGT theory, we consider the action in the following form:

$$S = \int L d^4 x = \int d^4 x \left\{ \frac{M_{Pl}^2}{2} \sqrt{-g} \left[ R + 2 m_g^2 U(g, \mathrm{K}) \right] + L_M \right\} \qquad (1)$$

Here, $M_{Pl}$ is the reduced Planck mass, $g$ is a metric determinant, $R$ is a Ricci scalar, and $m_g$ is a bare graviton mass parameter [11]. The Lagrangian $L_M$ represents the matter fields where the associated stress-energy tensor is defined by:

$$T_{\mu\nu} = 2(-g)^{-1/2} dL_M / dg_{\mu\nu} \tag{2}$$

Note that throughout this paper, we use natural units i.e. $h = c = 1$. The functional $U(g, K)$ is a ghost-free potential which can be expanded into the following sequence:

$$U(g, K) = U_2 + \alpha_3 U_3 + \alpha_4 U_4 \tag{3}$$

with $\alpha_i$, $i = 3, 4$ are dimensionless arbitrary parameters and

$$U_2(g, K) = \frac{1}{2!}\left([K]^2 - [K^2]\right) \tag{4}$$

$$U_3(g, K) = \frac{1}{3!}\left([K]^3 - 3[K][K^2] + 2[K^3]\right) \tag{5}$$

$$U_4(g, K) = \frac{1}{4!}\left([K]^4 - 6[K^2][K]^2 + 8[K^3][K] + 3[K^2]^2 - 6[K^4]\right) \tag{6}$$

where

$$K^\mu_\nu = \delta^\mu_\nu - \sqrt{g^{\mu\sigma} f_{\sigma\nu}} \tag{7}$$

$$f_{\mu\nu} = \partial_\mu \phi_a \partial_\nu \phi^a \tag{8}$$

and $[K] = K^\mu_\mu$. The symbol $\phi^a$ denotes the Stuckelberg real scalar field [11], while $g_{\mu\nu}$ and $f_{\mu\nu}$ define dynamical and fiducial metrics, respectively. The introduction of the stuckelberg field is aimed to restore the diffeomorphism invariance of the theory [3]. Explicit expressions of Eq. (4) – (6) are given in the appendix.

For our cosmological application purpose of the associated dRGT theory, we consider the FLRW universe. For this case, the general expression of the corresponding dynamical and fiducial metrics are given as follows:

$$g_{\mu\nu} = \text{diag}\left[-N^2, a^2(1-kr^2)^{-1}, a^2 r^2, a^2 r^2 \sin^2\theta\right] \tag{9}$$

$$f_{\mu\nu} = \text{diag}\left[-\dot{f}^2, b^2(1-kr^2)^{-1}, b^2 r^2, b^2 r^2 \sin^2\theta\right] \tag{10}$$

where the $N^2 > 0$ denotes the lapse function of the dynamical metric related to the coordinate-time interval $(dt)$ and proper-time $(d\tau)$ of a normal (Eulerian) observer i.e. $d\tau = N dt$ [13-15], and acting like a gauge function [14]. Geometrically, the proper-time interval is measured by an observer following the geodesic worldline with its unit timelike vector is perpendicular to the hypersurface of the foliated FLRW spacetime with constant coordinate-time [15]. The $a$ and $b$ symbols are the scale factors for the dynamical and fiducial metrics, respectively. The function $f$ is the stuckelberg scalar function $\phi^0 = f$ such that $\partial \phi^0 / \partial t = \dot{f}$ whereas $\phi^i = x^i$. The $k$ parameter is the spacetime curvature for the open $(k < 0)$, flat $(k = 0)$ and close $(k > 0)$ universe. Note that there are four quantities s that can be considered to vary the corresponding action i.e. $a$, $b$, $f$, and $N$.

Inserting metrics (9) and (10) into the corresponding dRGT action (1) yields the following effective Lagrangian:

$$L_{eff} = -\frac{3\dot{a}^2 a}{N} + 3kNa + m_g^2 Na^3 \left\{ 3\left(1-\frac{\dot{f}}{N}\right)\left(1-\frac{b}{a}\right) + 3\left(1-\frac{b}{a}\right)^2 \right.$$
$$\left. + \alpha_3 \left[ 3\left(1-\frac{\dot{f}}{N}\right)\left(1-\frac{b}{a}\right)^2 + \left(1-\frac{b}{a}\right)^3 \right] + \alpha_4 \left(1-\frac{\dot{f}}{N}\right)\left(1-\frac{b}{a}\right)^3 \right\} + \frac{1}{M_{Pl}^2} L_M \quad (11)$$

where the explicit form of $[K^n]$ in Eq. (4) – (6) for the dynamical and fiducial metrics of this FLRW case are given by: $[K^n] = (1-\dot{f}/N)^n + 3(1-b/a)^n$, with $n = 1, 2, 3, 4$. Varying the corresponding action with respect to $N$ and by considering Eq. (2) and (3) leads to the following dynamical equation:

$$\frac{\dot{a}^2}{a^2 N^2} + \frac{k}{a^2} = \frac{\rho}{3M_{Pl}^2} - \frac{m_g^2 X}{3} \quad (12)$$

with:

$$X = 3\left(1-\frac{b}{a}\right) + (1+\alpha_3)\left(1-\frac{b}{a}\right)^2 + (\alpha_3 + \alpha_4)\left(1-\frac{b}{a}\right)^3 \quad (13)$$

Here, we consider the bare graviton mass $m_g > 0$ as a free parameter which is similar to what was considered in ref. [8]. This consideration is in contrast with our previous report [11], in which the bare mass of graviton is considered to be bound rather than a free parameter.

Let us consider Eq. (12). There are two ways for handling it, namely either by solving the differential equation of scale factor $a(t)$ and fixing the coordinate-time lapse function $N(t)$, or by fixing $a(t)$ to get an explicit form of $N(t)$. In the previous studies e.g. summarized in [15], the lapse function was never specifically discussed, since they were focusing on the dynamics of the scale factor $a$ and simply considered the normal observer with trivial gauge choice $N = 1$ on the discussion. In this paper, we focus on the second perspective for a reason that will be explained in the following discussion.

For our purpose, we can also rewrite Eq. (12) in terms of the explicit form of lapse function as follows:

$$N^2 = \frac{3M_{Pl}^2}{\left[ \rho - M_{Pl}^2 m_g^2 X - 3M_{Pl}^2 k a^{-2} \right]} \frac{\dot{a}^2}{a^2} \quad (14)$$

Expressing explicitly the non-trivial coordinate-time lapse function $N$ as given in Eq. (14) can be considered as a new perspective to study the different aspect of the cosmological application of dRGT theory. It should be emphasized that from the Eq. (14) point of view, the corresponding form of the scale factor is assumed can be constructed phenomenologically based on observational cosmological data instead of solving Eq. (12) with unspecified lapse function.

On the other hand, varying the corresponding action with respect to $a$ yields the following dynamical equation:

$$\ddot{a} - \frac{\dot{a}^2}{a} - \frac{\dot{a}\dot{N}}{N} - \frac{N^2 k}{a} + \frac{aN^2 m_g^2 Y}{2} = -\frac{aN^2}{2M_{Pl}^2}(\rho + p) \tag{15}$$

with:

$$Y = \left(\frac{b}{a} - \frac{\dot{f}}{N}\right)\left[1 + 2(1+\alpha_3)\left(1 - \frac{b}{a}\right) + (\alpha_3 + \alpha_4)\left(1 - \frac{b}{a}\right)^2\right] \tag{16}$$

The stress-energy tensor (2) that has been commonly considered in the most related reports were given in the form of perfect fluid

$$T_{\mu\nu} = \text{diag}\left[N^2 \rho, a^2 p, a^2 p r^2, a^2 p r^2 \sin^2 \theta\right] \tag{17}$$

where $\rho = \rho_m + \rho_r$ denotes the total energy density of matter $(\rho_m)$ and radiation $(\rho_r)$, while $p = p_m + p_r$ represents the corresponding pressures which is given by the state equation: $p_{m(r)} = w_{m(r)}\rho_{m(r)}$ with $w_m = 0$ and $w_r = 1/3$ [16]. It is worth to note that from the conservation of stress-energy tensor (17), that is $\nabla_\mu T^{\mu\nu} = 0$, give us the dynamical equation of both $\rho_{m(r)}$

$$\dot{\rho}_{m(r)} + 3\frac{\dot{a}}{a}\left(\rho_{m(r)} + p_{m(r)}\right) = 0 \tag{18}$$

Such that the solutions of Eq. (18) leads to following total density solution:

$$\rho = \rho_{m,0} a^{-3} + \rho_{r,0} a^{-4} \tag{19}$$

Clearly for an expanding universe model with $\lim_{t \to \infty} a \to \infty$ implies that $\lim_{t \to \infty} \rho = 0$.

Next, by differentiating the lapse function (14) with respect to time, inserting the result into Eq. (15) and considering Eq. (18) we found that the following condition should be satisfied:

$$Y - \frac{a\dot{X}}{3\dot{a}} = 0 \tag{20}$$

Calculating the $\dot{X}$ from Eq. (13) and inserting the result along with Eq. (16) into Eq, (20) yield the following condition:

$$\left(\frac{\dot{b}}{\dot{a}} - \frac{\dot{f}}{N}\right)\left[1 + 2(1+\alpha_3)\left(1 - \frac{b}{a}\right) + (\alpha_3 + \alpha_4)\left(1 - \frac{b}{a}\right)^2\right] = 0 \tag{21}$$

Interestingly, the consistency condition between the results on the variation with respect to $b$ and $f$ lead also to the condition (21). Detailed derivations are given in the appendix.

Here, from the results reported in ref. [11], we only consider the following branch of condition (21):

$$1+2(1+\alpha_3)\left(1-\frac{b}{a}\right)+(\alpha_3+\alpha_4)\left(1-\frac{b}{a}\right)^2 = 0 \qquad (22)$$

From this condition one finds:

$$b_\pm = \left[\frac{2\alpha_3+\alpha_4+1\pm\sqrt{\alpha_3^2+\alpha_3-\alpha_4+1}}{\alpha_3+\alpha_4}\right]a \qquad (23)$$

It is obvious that the Eq. (23) holds the conditions: $\alpha_4 \neq -\alpha_3$ and $\alpha_4 \neq -3(\alpha_3+1)$, where both conditions exclude the singular case and ensure that $b_\pm \neq 0$, respectively. Inserting Eq. (23) into Eq. (13) leads to:

$$X \to X_\pm(\alpha_3,\alpha_4) = \frac{(1+\alpha_3)(2+\alpha_3+2\alpha_3^2-3\alpha_4)\pm 2\Gamma^{3/2}}{(\alpha_3+\alpha_4)^2} \qquad (24)$$

where

$$\Gamma = 1+\alpha_3+\alpha_3^2-\alpha_4 \qquad (25)$$

Clearly from Eq. (24) that the Eq. (12) describes a Friedmann-Lemaitre equation with the graviton-mass related cosmological-constant-like term. For the case of $X_\pm < 0$ ($X_\pm > 0$) the corresponding term plays as positive (negative) cosmological-constant-like term responsible for an accelerated (a decelerated) universe expansion. Clearly, the case of $X_\pm > 0$ does not leads Eq. (12) to constitute a self-accelerating universe which is the main cosmological application of the related dRGT theory.

Based on this result the Eq. (14) can now be written as:

$$N^2 = \frac{3M_{Pl}^2}{\left[\rho - 3M_{Pl}^2 ka^{-2} - M_{Pl}^2 m_g^2 X_\pm(\alpha_3,\alpha_4)\right]}\frac{\dot{a}^2}{a^2} \qquad (26)$$

It is readily seen that the behavior of $N$ given in Eq. (26) for an expanding universe, where $\lim_{t\to\infty} a \to \infty$ and $\lim_{t\to\infty} \rho = 0$, depends on the sign of $X_\pm$. This fact clearly demonstrates that the characteristic of $N$ in the corresponding parameter space $(\alpha_3,\alpha_4)$ can exhibit unphysical behavior for $X_\pm > 0$, which is associated with the negative graviton-mass related cosmological-constant-like term. Therefore, the related allowed parameter space cannot entirely be related to the accelerated expansion of the universe. Analysis of the characteristics of $X_\pm$ in the parameter space $(\alpha_3,\alpha_4)$ has been discussed in details in refs. [8, 11].

To solve this unphysical problem of $N$, we can consider a simple modification on the stress-energy tensor (17) namely by introducing a new parameter $\rho_0$ such that $\lim_{t\to\infty}\rho = \rho_0$, whereas to preserve the dynamical equation of matter-radiation density (18) it should also satisfy a condition of $\lim_{t \to \infty} p = -\rho_0$. For this purpose, we consider the following simple transformations:

$$\rho \to \rho + \rho_0 \tag{27}$$

and

$$p = w\rho \to p = w\rho - \rho_0 \tag{28}$$

such that:

$$T_{\mu\nu} = \text{diag}\left[ N^2(\rho + \rho_0), a^2(w\rho - \rho_0), a^2(w\rho - \rho_0)r^2, a^2(w\rho - \rho_0)r^2 \sin^2\theta \right] \tag{29}$$

Clearly, based on these transformations, the stress-energy tensor (29) at $t \to \infty$ becomes:

$$\lim_{t \to \infty} T_{\mu\nu} = \text{diag}\left[ N^2 \rho_0, -a^2 \rho_0, -a^2 \rho_0 r^2, -a^2 \rho_0 r^2 \sin^2\theta \right] \tag{30}$$

This condition demonstrates that the newly introduced parameter $\rho_0$ implies negative pressure that leads to the accelerated expansion of the universe. In terms of the conventional model for dark-energy [1] without massive graviton, this parameter plays a similar role with the cosmological-constant [8] in the corresponding Friedmann-Lemaitre equation. We prefer to consider the modification of perfect fluid stress-energy tensor rather than the conventional cosmological-constant $\Lambda$ in our formalism which is commonly interpreted as dark energy instead of ordinary energy. However, both concepts are equivalent. To distinct with the graviton-mass related cosmological-constant-like term, we call this proposed parameter as a "cosmological background density" parameter.

Based on transformation (27) and (28) the Eq. (26) is now becomes:

$$N^2 = \frac{3M_{Pl}^2}{\left[ \rho + \rho_0 - 3M_{Pl}^2 k a^{-2} - M_{Pl}^2 m_g^2 X_{\pm}(\alpha_3, \alpha_4) \right]} \frac{\dot{a}^2}{a^2} \tag{31}$$

Clearly, by introducing the related parameter one can remove the aforementioned unphysical problem and restore the positive value of the effective cosmological-constant-like term that consists of cosmological background density and graviton-mass related cosmological-constant-like terms given in the right-hand side of Eq. (12). In other words, the related modified perfect fluid stress-energy tensor leads the model only to admit accelerated expanding universe model. Note that for the case of $X_{\pm} < 0$ the model is already described an accelerated expansion universe model with physically acceptable lapse function, such that one can set $\rho_0 = 0$. However, $\rho_0 \neq 0$ is still required when the value of the actual graviton-mass value is observed to be insufficient to provide a cosmological-constant-like term in the allowed parameter space that leads to the presently observed universe expansion. In the following section, we discuss a simple example.

## 3. Flat FLRW universe with exponential scale factor

As discusses previously, one can consider that the scale factor $a$ can be constructed from the observational cosmological data. Based on this, we could simply consider a case of a flat universe with $k = 0$, which is relevant to the fact that our present universe is observed to be nearly flat [17], and a simple scale factor in terms of coordinate-time $(t)$ describing the accelerated era namely:

$$a = a_0 \exp(Ht) \tag{32}$$

in a spherical symmetric coordinates, the dynamical and fiducial metrics have the forms:

$$g_{\mu\nu} = \text{diag}\left[-N^2,\ a_0^2 e^{2Ht},\ a_0^2 e^{2Ht} r^2,\ a_0^2 e^{2Ht} r^2 \sin^2\theta\right] \tag{33}$$

$$f_{\mu\nu} = \text{diag}\left[-\dot{f}^2,\ b_0^2 e^{2Ht},\ b_0^2 e^{2Ht} r^2,\ b_0^2 e^{2Ht} r^2 \sin^2\theta\right] \tag{34}$$

where the parameter $H$ represents the constant Hubble rate. From Eq. (14), (19) and (32), the expression for lapse function (31) for this case is given by:

$$N^2 = \frac{3H^2 M_{Pl}^2}{a_0 \rho_{m,0} e^{-3Ht} + a_0 \rho_{r,0} a^{-4Ht} - M_{Pl}^2 m_g^2 X_\pm(\alpha_3, \alpha_4)} \tag{35}$$

For the sake of simplicity, the Eq. (35) can be transformed into a normalized form:

$$N^2 = \frac{1}{\tilde{\rho}_{m,0} e^{-3Ht} + \tilde{\rho}_{r,0} a^{-4Ht} - \tilde{X}_\pm(\alpha_3, \alpha_4)} \tag{36}$$

where the normalized symbols $\tilde{\rho}$ and $\tilde{X}_\pm$ represent:

$$\tilde{\rho}_{m(r),0} \equiv \frac{a_0 \rho_{m(r),0}}{3H^2 M_{Pl}^2} \quad \text{and} \quad \tilde{X}_\pm \equiv \frac{m_g^2 X_\pm}{3H^2} \tag{37}$$

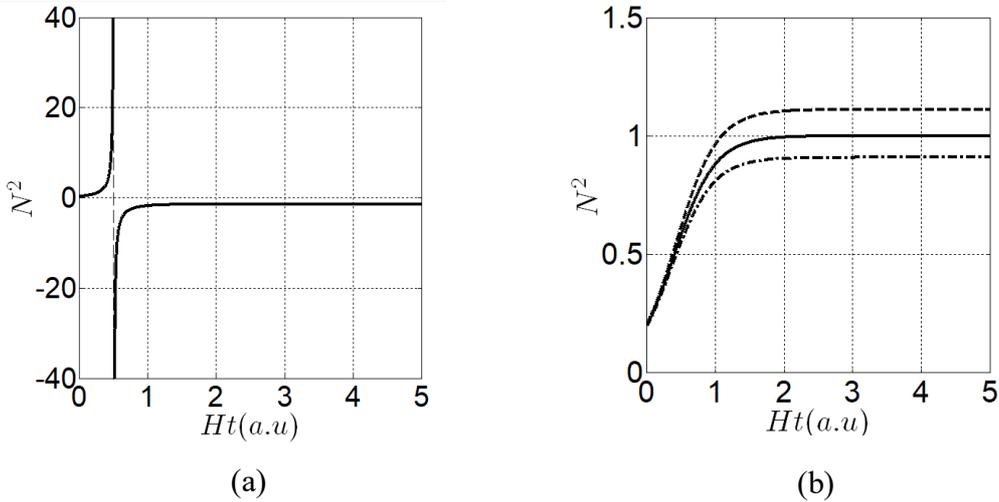

(a)                    (b)

Fig. 1 Illustration of Eq. (38) plot for the case of (a) $\tilde{\rho}_0 \to 0$ with $\tilde{X}_\pm = 0.7$, and (b) for $\tilde{\rho}_0 = 1.7$ with $\tilde{X}_\pm = 0.6$ (dash-dotted-curve), $\tilde{X}_\pm = 0.7$ (solid-curve), and $\tilde{X}_\pm = 0.8$ (dash-curve).

Respectively, which are illustrated in Fig. 1a an example of $N$ showing the existence of singularity and negative value.

It is clear that the corresponding lapse function $N$ given by Eq. (31) in a normalized form becomes:

$$N^2 = \frac{1}{\tilde{\rho}_{m,0} e^{-3Ht} + \tilde{\rho}_{r,0} a^{-4Ht} + \tilde{\rho}_0 - \tilde{X}_{\pm}(\alpha_3, \alpha_4)} \tag{38}$$

where $\tilde{\rho}_{m(r),0}$ and $\tilde{X}_{\pm}$ are defined by Eq. (38). Similarly, the symbol $\tilde{\rho}_0 \equiv \rho_0 H^{-2} M_{Pl}^{-2}/3$. Under the following parameter condition:

$$\tilde{\rho}_0 > \tilde{X}_{\pm}(\alpha_3, \alpha_4) \tag{39}$$

for the $X_{\pm} > 0$ condition, it is readily seen that in general, the expression (38) admits a physically acceptable behavior lapse function for all allowed regions in the parameter space $(\alpha_3, \alpha_4)$ which is related to a self-accelerating universe model. An example is given in Fig. 1b to illustrate the dynamics of Eq. (38) i.e. $X_{\pm} > 0$. It is clearly shown that the related lapse function demonstrates saturated characteristic at $t \to \infty$ which indicates a constant relation between the coordinate-time and proper-time for normal observer.

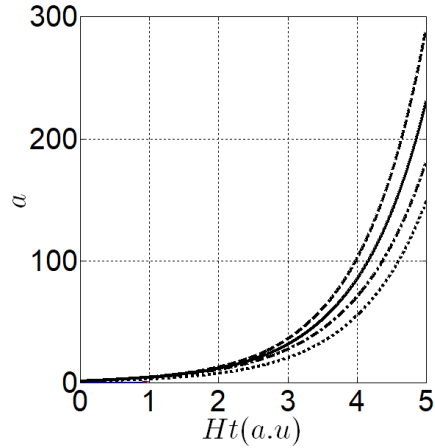

Fig. 2  Illustration of the scale factor plot in terms of coordinate-time (dotted-curve) and proper-time for the case of $\tilde{X}_{\pm} = 0.6$ (dash-dotted-curve), $\tilde{X}_{\pm} = 0.7$ (solid-curve), and $\tilde{X}_{\pm} = 0.8$ (dash-curve). The cosmological background density is set to $\tilde{\rho}_0 = 1.7$.

At this point, it is also interesting to compare the behavior of scale factor in terms of coordinate-time $(t)$ given by Eq. (32) with the scale factor in terms of proper-time $(\tau)$ which can be found by solving numerically Eq. (12) for $N = 1$. For this, we consider a dimensionless coordinate-time $t \to Ht$, where $H = \dot{a}/a$ is a constant, such that Eq. (12) reads:

$$\dot{a}^2 = \frac{1}{3H^2 M_{Pl}^2}\left(\rho_m + \rho_r + \rho_0 - M_{Pl}^2 m_g^2 X\right) a^2 = \left(\tilde{\rho}_m + \tilde{\rho}_r + \tilde{\rho}_0 - \tilde{X}_{\pm}\right) a^2 \tag{40}$$

for $N=1$ and $k=0$. Here, the definition of $\tilde{\rho}_{m,r,0}$ and $\tilde{X}_\pm$ are given by Eq. (37), while the differential equation of the normalized matter-radiation densities $\tilde{\rho}_{m(r)}$ in the dimensionless coordinate-time are given as follows:

$$\dot{\tilde{\rho}}_{m(r)} + 3\left(1 + w_{m(r)}\right)\tilde{\rho}_{m(r)} = 0 \tag{41}$$

with $w_m = 0$ and $w_r = 1/3$ [15].

Depicted in Fig. 2 are the scale factors with a fix $\tilde{\rho}_0$ in terms of proper-time $(N=1)$ for varying $\tilde{X}_\pm$, which are calculated from the Eq. (40) and (41) using standard Runge-Kutta method, and the related exponential scale factor in terms of the dimensionless coordinate-time $(Ht)$ variable given by Eq. (32) with $a_0 = 1$. For the numerical calculation we consider as the related initial conditions: $a(0) = 1$ and $\tilde{\rho}_{m(r)}(0) = 2$.

As expected, Fig. 2 shows that the variation of $X_\pm$ leads to the different proper-time scale factor, while the exponential coordinate-time scale factor is fixed. The connection between the scale factor for the coordinate-time $(N \neq 1)$ and proper-time shown in Fig. 2 is associated with the coordinate-time lapse function given by Fig. 1b. Note that these two different time variables are related through the relation $d\tau = Ndt$. Therefore, it is important to emphasize that the proper-time scale factor is observed by a normal observer, at which from the coordinate-time point of view is given by Eq. (32), where the related lapse function in the dynamical metric (33) is free from unphysical behavior in all parameter space.

## 4. Possible range for $\rho_0$ parameter

At this point, it is intriguing enough to further investigate the true physical meaning of the newly proposed $\rho_0$ parameter, where its value is in order of $M_{Pl}^2 m_g^2$. However, attempt to determine its present in the universe and to measure its value observationally is beyond the scope of our study. Nevertheless, we can make a prediction on the possible range of the value of this parameter.

Let us assume that the bare graviton mass is equal to the lowest Higuchi bound under specific gauge condition of $\dot{f}/N = 1$ [11] for the functions given in metrics (9) and (10) namely $m_g \geq \sqrt{2}H_0$, where $H_0 = 1.382 \times 10^{-33}\ eV$ is the Hubble constant of the present universe given in natural unit, such that $m_g \geq 1.954 \times 10^{-33}\ eV$. Since the reduced Planck mass in the natural unit is $M_{Pl} = 2.435 \times 10^{27}\ eV$, then $M_{Pl}^2 m_g^2 \geq 2.264 \times 10^{-11}\ eV^4$.

Now let us consider the following effective form:

$$\rho_0 - M_{Pl}^2 m_g^2 X_\pm = 3H_0^2 M_{Pl}^2 \Omega_{DE} \tag{42}$$

of the self-accelerating term on the left-hand side, where $\Omega_{DE}$ is defined as density parameter for dark-energy, such that:

$$\rho_0 = M_{Pl}^2 m_g^2 \left(\frac{3H_0^2 \Omega_{DE}}{m_g^2} + X_\pm\right) \tag{43}$$

which is automatically satisfies the condition (39) that ensures the physically acceptable lapse function $N$. By considering the present measured value $\Omega_{DE} \approx 0.7$ [17] leads to the following possible range for the $\rho_0$ parameter for the abovementioned chosen bare graviton mass:

$$\rho_0 \geq 2.264 \times 10^{-11} (1.050 + X_{\pm}) eV^4 \qquad (44)$$

which requires the following condition:

$$1.050 + X_{\pm} > 0 \qquad (45)$$

As an illustration, if the corresponding cosmological background density is nothing but the vacuum energy density of the present universe i.e. $\rho_0 = \rho_{vac} = 4.310 \times 10^{-11} eV^4$, then based on conditions (44) and (45) one finds $-1.050 < X_{\pm} \leq 0.854$. Note that the examples given by Fig. 1b and Fig. 2 are in this interval.

## 5. Conclusion

We have shown that the consideration of perfect fluid assumption on stress-energy tensor in dRGT theory has led to the existence of the unphysical condition for the coordinate-time lapse function of FLRW metrics in the parameter space. To solve this problem we propose the modification of the related perfect fluid form by introducing a new parameter $\rho_0$. This parameter is interpreted as a cosmological background density that induces negative pressure at $t \to \infty$ which dominates the negative cosmological-constant-like term induced by the graviton mass term. We have demonstrated that this proposal can restore the physically acceptable characteristic of the corresponding non-trivial coordinate-time lapse function in the all allowed parameter space for a self-accelerating universe model. It should be emphasized that our present report on the dynamics of the corresponding gauge-like lapse function can be considered to offer a different aspect on the cosmological application of dRGT theory, where it seems that the associated dRGT theory also implicitly suggests the existence of another unknown quantity in the universe other than the massive graviton which is represented by $\rho_0$ parameter in our formulation. The possible range of this parameter was also discussed.

## 6. Acknowledgement

The work of HA is partially funded by "Riset Inovasi ITB" 2017 and the Directorate of Research and Innovation, Bogor Agricultural University, under decree no. 186/IT3/PN/2016, while the work of BEG is supported by "Riset Inovasi ITB" 2017 and PDUPT Kemenristekdikti 2018-2019.

## Appendix

We discuss in this appendix the application of the effective Lagrangian approach to find the Friedmann-Lemaitre and constraint equations for the chosen metrics. Consider the action given by Eq. (1), and the dynamical and fiducial metrics given by Eq. (9) and (10), respectively. Inserting both metrics to $K_{\nu}^{\mu}$ lead to

$$K_0^0 = 1 - \frac{\dot{f}}{N}; \quad K_j^i = \left(1 - \frac{b}{a}\right)\delta_j^i; \quad K_0^i = K_i^0 = 0 \qquad (A1)$$

and yielding:

$$\mathrm{K} = \mathrm{K}_\mu^\mu = 1 - \frac{\dot{f}}{N} + 3\left(1 - \frac{b}{a}\right) \tag{A2}$$

$$U_2(g,f) = \frac{1}{2!}\left([\mathrm{K}]^2 - [\mathrm{K}^2]\right) = 3\left(1 - \frac{\dot{f}}{N}\right)\left(1 - \frac{b}{a}\right) + 3\left(1 - \frac{b}{a}\right)^2 \tag{A3}$$

$$U_3(g,f) = \frac{1}{3!}\left([\mathrm{K}]^3 - 3[\mathrm{K}][\mathrm{K}^2] + 2[\mathrm{K}^3]\right) = 3\left(1 - \frac{\dot{f}}{N}\right)\left(1 - \frac{b}{a}\right)^2 + \left(1 - \frac{b}{a}\right)^3 \tag{A4}$$

$$U_4(g,f) = \frac{1}{4!}\left([\mathrm{K}]^4 - 6[\mathrm{K}^2][\mathrm{K}]^2 + 8[\mathrm{K}^3][\mathrm{K}] + 3[\mathrm{K}^2]^2 - 6[\mathrm{K}^4]\right) = \left(1 - \frac{\dot{f}}{N}\right)\left(1 - \frac{b}{a}\right)^3 \tag{A5}$$

Based on the corresponding Eq. (A1), we found the expressions for each component of potential function $U(g,\mathrm{K})$ which are given by Eq. (A2) – (A5) such that:

$$\begin{aligned}U(g,f) = &\; 3\left(1 - \frac{\dot{f}}{N}\right)\left(1 - \frac{b}{a}\right) + 3\left(1 - \frac{b}{a}\right)^2 + \alpha_3\left[3\left(1 - \frac{\dot{f}}{N}\right)\left(1 - \frac{b}{a}\right)^2 + \left(1 - \frac{b}{a}\right)^3\right] \\ &+ \alpha_4\left(1 - \frac{\dot{f}}{N}\right)\left(1 - \frac{b}{a}\right)^3\end{aligned} \tag{A6}$$

Based on these results, the corresponding effective action is then given by:

$$S = M_{Pl}^2 r^2 \sin\theta \int d^4x\, L_{eff} \tag{A7}$$

with

$$L_{eff} = -\frac{3\dot{a}^2 a}{N} + 3kNa + m_g^2 Na^3 U(g,f) + \frac{1}{M_{Pl}^2} L_M \tag{A8}$$

is the related effective Lagrangian with $U(g,f)$ is given by Eq. (A6). Here, $L_M$ denotes the effective Lagrangian for matter which is related to the stress-energy tensor given by Eq. (15). The term $\sqrt{-g}$ term in the associated action (1) is given by: $\sqrt{-g} = Na^3 r^2 \sin\theta$.

First, varying (A7) with respect to $N$ leads to the following Euler-Lagrange equation:

$$\begin{aligned}\frac{\partial L}{\partial N} - \frac{d}{dt}\left(\frac{\partial L}{\partial \dot{N}}\right) = &\; \frac{3\dot{a}^2 a}{N^2} + 3ka + m_g^2 a^3 \left[3\left(1 - \frac{b}{a}\right) + \left(1 - \frac{b}{a}\right)^2 \right. \\ &\left. + \alpha_3\left(1 - \frac{b}{a}\right)^2 + \alpha_3\left(1 - \frac{b}{a}\right)^3 + \alpha_4\left(1 - \frac{b}{a}\right)^3\right] - \frac{a^3}{M_{Pl}^2}\rho = 0\end{aligned} \tag{A9}$$

Next, varying (A7) with respect to $a$ leads to the following Euler-Lagrange equation:

$$\frac{\partial L}{\partial a} - \frac{d}{dt}\left(\frac{\partial L}{\partial \dot{a}}\right) = -\frac{3\dot{a}^2}{N} + 3kN + m_g^2 N \left\{ 3(2a-b)\left(2a - \frac{a\dot{f}}{N} - b\right) + 3a(a-b)\left(2 - \frac{\dot{f}}{N}\right) \right.$$
$$+ \alpha_3 \left[ 2(a-b)\left(4a - \frac{3a\dot{f}}{N} - b\right) + (a-b)^2\left(4 - \frac{3\dot{f}}{N}\right) \right]$$
$$\left. 3\alpha_4 (a-b)^2 \left(1 - \frac{\dot{f}}{N}\right) \right\} + \frac{3Na^2}{M_{Pl}^2}P + \frac{6\ddot{a}a}{N} + \frac{6\dot{a}^2}{N} - \frac{6\dot{a}^2 a\dot{N}}{N^2} = 0$$
(A10)

After a straightforward manipulation Eq. (A10) can then be recast into Eq. (15).

The next step is varying the effective action (A7) with respect to $b$ and $f$. First, we vary with respect to $b$ that leads to the following Euler-Lagrange equation:

$$\frac{\partial Na^3 U}{\partial b} - \frac{d}{dt}\left(\frac{\partial Na^3 U}{\partial \dot{b}}\right) = -3a^2 \left\{ (N - \dot{f})\left[1 + 2\alpha_3\left(1 - \frac{b}{a}\right) + \alpha_4\left(1 - \frac{b}{a}\right)^2\right] \right.$$
$$\left. + 2N\left(1 - \frac{b}{a}\right) + \alpha_3 N\left(1 - \frac{b}{a}\right)^2 \right\} = 0$$
(A11)

yielding

$$\left(1 - \frac{\dot{f}}{N}\right)\left[1 + 2\alpha_3\left(1 - \frac{b}{a}\right) + \alpha_4\left(1 - \frac{b}{a}\right)^2\right] + 2\left(1 - \frac{b}{a}\right) + \alpha_3\left(1 - \frac{b}{a}\right)^2 = 0 \quad (A12)$$

Variation of action (A7) with respect to $f$ yielding the following Euler-Lagrange equation:

$$\frac{\partial Na^3 U}{\partial f} - \frac{d}{dt}\left(\frac{\partial Na^3 U}{\partial \dot{f}}\right) = -3a^2 \left\{ (\dot{a} - \dot{b})\left[1 + 2\alpha_3\left(1 - \frac{b}{a}\right) + \alpha_4\left(1 - \frac{b}{a}\right)^2\right] \right.$$
$$\left. + 2\dot{a}\left(1 - \frac{b}{a}\right) + \alpha_3 \dot{a}\left(1 - \frac{b}{a}\right)^2 \right\} = 0$$
(A13)

such that

$$\left(1 - \frac{\dot{b}}{\dot{a}}\right)\left[1 + 2\alpha_3\left(1 - \frac{b}{a}\right) + \alpha_4\left(1 - \frac{b}{a}\right)^2\right] + 2\left(1 - \frac{b}{a}\right) + \alpha_3\left(1 - \frac{b}{a}\right)^2 = 0 \quad (A14)$$

Equating Eq. (A12) and Eq. (A14) leads to the following condition:

$$\left(\frac{\dot{b}}{\dot{a}} - \frac{\dot{f}}{N}\right)\left[1 + 2(1 + \alpha_3)\left(1 - \frac{b}{a}\right) + (\alpha_3 + \alpha_4)\left(1 - \frac{b}{a}\right)^2\right] = 0 \quad (A15)$$

which is resembling the condition given by Eq. (21).